\documentclass[aps,prl,groupedaddress,twocolumn,showpacs,preprintnumbers,amsmath,amssymb]{revtex4}
\usepackage{graphicx}% Include figure files
\usepackage{dcolumn}% Align table columns on decimal point

\begin{document}

\title{Time-Resolved Spin Torque Switching and Enhanced Damping in Py/Cu/Py Spin-Valve Nanopillars}

\author{N. C. Emley}
\author{I. N. Krivorotov}
\author{A. G. F. Garcia}
\author{O. Ozatay}
\author{J. C. Sankey}
\author{D. C. Ralph}
\author{R. A. Buhrman}

% \email{rab8@cornell.edu}
\affiliation{Cornell University, Ithaca, NY 14853-2501}


\date{\today}

\begin{abstract}
We report time-resolved measurements of current-induced reversal of a free magnetic layer in Py/Cu/Py elliptical
nanopillars at temperatures $T$ = 4.2 K to 160 K.  Comparison of the data to Landau-Lifshitz-Gilbert macrospin
simulations of the free layer switching yields numerical values for the spin torque and the Gilbert damping
parameters as functions of $T$.  The damping is strongly $T$-dependent, which we attribute to the
antiferromagnetic pinning behavior of a thin permalloy oxide layer around the perimeter of the free layer.  This
adventitious antiferromagnetic pinning layer can have a major impact on spin torque phenomena.
\end{abstract}

\maketitle

Experiments~\cite{katine-2000PRL,urazhdin-2003PRL,kiselev-2003Nature,grollier-2000APL} have shown that a
spin-polarized current passed through a nanomagnet can excite a dynamic response as the result of a spin torque
applied by the conduction electrons~\cite{slonczewski-1996JMMM,berger-1996PRB}.  The potential for technological
impact of this spin transfer (ST) effect has inspired research in DC current-induced microwave
oscillations~\cite{kiselev-2003Nature,rippard-2005PRL} and hysteretic
switching~\cite{katine-2000PRL,urazhdin-2003PRL,grollier-2000APL} in current perpendicular to the plane (CPP)
nanopillars and nanoconstrictions. Typically, ST switching data is obtained through the use of slow current ramp
rates ($\sim$1 mA/s), but fast pulses ($\sim$10$^{10}$ mA/s) access the regime where thermal activation of the
moment over a current-dependent barrier~\cite{li-zhang-2004PRB,krivorotov-2004PRL} does not play a major role in
the switching process.  This spin torque-driven regime~\cite{koch-2004PRL} is advantageous for the quantitative
examination of the spin torque parameters due to the computational accessibility of numerically integrating the
Landau-Lifshitz-Gilbert (LLG) equation for short durations.

Here we report time-resolved measurements of the spin torque-driven switching event in Cu 100/Py 20/Cu 6/Py 2/Cu
2/Pt 30 (in nm, Py = Ni$_{81}$Fe$_{19}$) CPP spin-valve nanopillar structures at bath temperatures $T$ = 4.2 K
to 160 K. We compare our experimental results with LLG simulations in the macrospin approximation and find good
agreement between simulation and measurement. This both confirms the applicability of the macrospin
approximation in the spin torque-driven regime and facilitates the quantitative determination of $T$-dependent
spin torque and magnetic damping parameters.  At higher $T$ we find that the strength of the spin torque exerted
per unit current is in reasonable numerical accord with recent model calculations, and that the damping
parameter $\alpha_{0}$ for the nanomagnet excitations is both anomalously high, as suggested by previous pulsed
current measurements~\cite{braganca-2005APL}, and $T$-dependent. The strong $T$ variation of $\alpha_{0}$, in
conjunction with anomalous behavior of the nanomagnet switching fields $H_{S,i}$($T$) in some devices, points to
the presence of an adventitious antiferromagnetic oxide layer around the perimeter of the nanomagnet that has a
major effect on the nanomagnet dynamics driven by a spin torque.

\begin{figure}
\includegraphics[width=8.5cm]{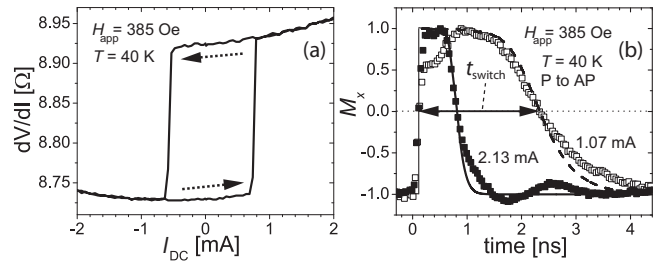}
\caption{(a) Slow ramp rate spin torque switching of a nanomagnet as measured by the GMR effect for sample 1, a
60$\times$190 nm ellipse, at $T$ = 40 K and $H_{\textrm{app}}$ = 385 Oe, which opposes the dipole field so that
$H_{\textrm{app}} + H_{\textrm{dip}} \approx 0$. Arrows indicate the scan direction. (b) Pulsed spin torque P to
AP switching measured for the same sample at $I$ = 1.07 mA ($\square$) and 2.13 mA ($\blacksquare$). The data
(symbols) have been normalized to $M_{x}$ = $\pm$1 for simple comparison with the simulated macrospin switching
(lines). Pulse shape distortions are due to the setup (see ref. [13]).}
\end{figure}

The nanopillar devices employed in this study were fabricated using a process described
elsewhere~\cite{emley-2004APL}. A slow ramp rate ST scan is shown in Fig. 1(a) for sample 1, a 60$\times$190 nm
ellipse. In Fig. 1(b) we show parallel (P) to anti-parallel (AP) switching events for sample 1, averaged over
10,000 switches, taken at pulsed current amplitudes $I$ = 1.07 mA and 2.13 mA at $T$ = 40 K as open and solid
squares, respectively.  The measured signal is a time-resolved voltage drop $|I\cdot\Delta R|$ from the giant
magnetoresistance (GMR) of the sample as the free layer switches from P to AP orientation, where $\Delta R
\equiv R_{x}$(AP)$ - R_{x}$(P) and $R_{x}$ is the 4-point device resistance.  The data have been normalized to
$M_{x} = +/- 1$ (minimum resistance / maximum resistance) for simple comparison with simulated switching events,
described below, which are shown as solid and dashed lines. The abrupt ($\sim$200 ps) jump from $M_{x}$ = -1 to
1 at time = 0 is not a switching event but is simply the rising edge of the current pulse. The more gradual
transition between P ($M_{x}$ = 1) and AP (return to $M_{x}$ = -1) is the envelope coming from averaging over
thousands of individual switching events, each of which follows a trajectory determined by initial conditions
that are randomized by the stochastic thermal fluctuations of the free layer. We define the switching time
$t_{switch}$ as the time elapsed between 50\% of the signal rise and 50\% of the signal drop as indicated in
Fig. 1(b)~\cite{jitter}.

To obtain a quantitative understanding of the ST switching, we have simulated the nanomagnet dynamics by
numerical integration of the LLG equation in the macrospin approximation with the inclusion of a
Slonczewski-type spin torque term.
\begin{eqnarray*}
\frac{d\hat{m}}{dt} = \gamma[\hat{m} \times (\vec{H}_{\textrm{eff}} + \vec{H}_{\textrm{Lang}}(T^{\prime})) -
\alpha(\theta) \hat{m} \times (\hat{m} \times (\vec{H}_{\textrm{eff}}\\ + \vec{H}_{\textrm{Lang}}(T^{\prime})))
- \frac{I\hbar g(\theta)}{e M_{s}(T^{\prime})(area\cdot d)\sin\theta} \hat{m}\times\hat{p}\times\hat{m}]
\end{eqnarray*}
\newline Here $\gamma$ is the gyromagnetic ratio, $\hat{m}$ is the unit directional vector of the
free layer macrospin, $\hat{p}$ is the spin polarization axis, $\theta$ is the in-plane angle between them,
$g(\theta)$ is the spin torque function, $M_{s}(T)$ is the free layer magnetization, as measured separately for
a continuous 2 nm Py film in a Cu/Py/Cu trilayer that was exposed to the same heat treatments as the
nanopillars, $d$ is the nanomagnet thickness, $area = \frac{\pi}{4}ab$ is its lateral area with dimensions $a$
and $b$ that are estimated by OOMMF micromagnetic simulations~\cite{OOMMF} (see below), and
$\vec{H}_{\textrm{eff}}$ is the sum of external $\vec{H}_{\textrm{ext}}$, in-plane anisotropy
$\vec{H}_{\textrm{K}}$, and out-of-plane anisotropy $\vec{H}_{\perp}$ fields. $\vec{H}_{\textrm{ext}}$ is the
sum of the magnetostatic dipole field from the fixed layer $\vec{H}_{\textrm{dip}}$ and the applied field
$\vec{H}_{\textrm{app}}$ from the electromagnet, which is adjusted to compensate for $\vec{H}_{\textrm{dip}}$ so
$\vec{H}_{\textrm{ext}} \approx 0$.

The initial conditions of the simulation were set by $\theta_{i}$ = $\theta_{0}$ + $\theta_{\textrm{mis}}$ +
$\theta_{\textrm{rand}}(T)$, where $\theta_{0}$ = 0$^{\circ}$ for P to AP and 180$^{\circ}$ for AP to P
switching and $\theta_{\textrm{mis}}$ represents any systematic angular misalignment between free and fixed
layer moments due to the setup and was generally set to 0.  The random angle $\theta_{\textrm{rand}}(T)$ is
treated as a Gaussian with a standard deviation $\sqrt{k_{B}T/2E_{0}(T)}$ where $E_{0}(T)=E_{0}(4.2 \textrm{
K})[M_{s}(T)/M_{s}(4.2 \textrm{ K})]^{2}$ is the uniaxial anisotropy energy. $E_{0}$(4.2 K), $a$, and $b$ are
estimated from $T$ = 0, 2D OOMMF simulations of Py elliptical disks having $H_{\textrm{K}}$ and $\Delta R$
values similar to those measured at 4.2 K. The lateral area is estimated to 12\% uncertainty with this method,
nearly a factor of 2 better than the inherent shape variation among otherwise identical elliptical patterns due
to lithographic fluctuations.  Ohmic heating effects during the current pulse are taken into account by locally
raising the temperature of the device to $T^{\prime}=\sqrt{T^{2} + 10.23(\textrm{K/mV})^{2}(R_{x}(T)\cdot
I)^{2}}$~\cite{krivorotov-2004PRL}. A Langevin field $\vec{H}_{\textrm{Lang}}(T^{\prime})$ accounts for thermal
fluctuations during the dynamic trajectory, fluctuating randomly in 3-dimensions with a standard deviation
$\sqrt{2\alpha_{0}k_{B}T^{\prime}\mu_{0}/\gamma M_{s}(T^{\prime})(area\cdot d)\Delta t}$ where $\Delta t$ = 1 ps
is the time step~\cite{li-zhang-2004PRB,koch-2004PRL,russek-2005PRB}. Gilbert damping is assigned an angular
dependence $\alpha(\theta) = \alpha_{0}[1 - \nu \sin^{2}\theta/(1 - \nu^{2} \cos^{2}\theta)]$, where $\nu$ =
0.33 for Py/Cu/Py nanopillars~\cite{tserkovnyak-2003PRB}, but the addition of this angle-dependent damping term
had only a small effect on the simulation results.

\begin{figure}
\includegraphics[width=8.5cm]{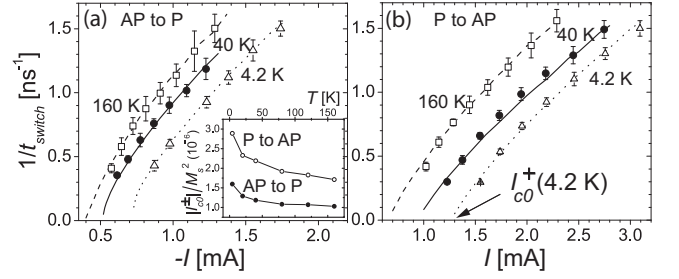}
\caption{Measured (symbols) and simulated (lines) 1/$t_{switch}$ versus $I$ for sample 2, an 80$\times$180 nm
ellipse, at $T$ = 160 K ($\square$), 40 K ($\bullet$), and 4.2 K ($\vartriangle$) for (a) AP to P and (b) P to
AP switching. Simulations to 1/$t_{switch}$ $\approx$ 0 yield estimates of the intercepts $I_{c0}^{\pm}(T)$. (a)
inset: $I_{c0}^{\pm}(T)/M_{s}^{2}(T)$.}
\end{figure}

The spin torque function is approximated by $g(\theta) = A \sin\theta/(1 + B \cos\theta)$ where $A$ and $B$ are
phenomenological parameters~\cite{braganca-2005APL,slonczewski-2002JMMM,xiao-2004PRB}. In our simulations we use
$\alpha_{0}$, $A$, and $B$ as $T$-dependent fitting parameters to match the simulated with the measured values
of 1/$t_{switch}$ versus $I$ for each $T$, where we allow $\alpha_{0}$ to be different for the two switching
directions.  In Fig. 1(b) we plot the average of 2000 simulated P to AP switching events at $T$ = 40 K alongside
the normalized data for sample 1 with the best fit simulation yielding $A$ = 0.5, $B$ = 0.11, and $\alpha_{0}$ =
0.048.  Since the current step in the simulation turns on instantaneously, an average pulse half rise time of
112 ps, measured from data such as those in Fig. 1(b), has been added to all simulated $t_{switch}$.  We plot
measured 1/$t_{switch}$ versus $I$ for AP to P and P to AP switching at $T$ = 160 K, 40 K, and 4.2 K for sample
2, an 80$\times$180 nm ellipse, together with best fit simulations, all of which are averages over 2000 events,
in Fig. 2(a) and 2(b), respectively. Simulations out to long switching times (1/$t_{switch} < 0.1$ ns$^{-1}$)
allow for good estimates of the 1/$t_{switch} \rightarrow 0$ intercepts $I_{c0}^{\pm}(T)$, which are the
critical currents (+ = P to AP) defining the onset of spin torque-driven switching. These should depend on the
spin torque and damping parameters as $I_{c0}^{\pm}(T) \propto \alpha_{0}M_{s}^{2}(T)$~\cite{koch-2004PRL}. A
striking result from these measurements is the strong $T$-dependence of $I_{c0}^{\pm}(T)/M_{s}^{2}(T)$ (Fig.
2(a) inset), which varies by more than 60\% over the entire $T$ range, where the upturns at low $T$ indicate a
strong dependence of damping, spin torque, or both.

In Fig. 3 the best fit values for $\alpha_{0}$, $A$, and $B$ (assuming $\theta_{\textrm{mis}}$ = 0$^{\circ}$)
are plotted as functions of $T$ for sample 2. Uncertainties in the fit parameters, $\Delta\alpha_{0}$ = 0.0035,
$\Delta A$ = 0.025, and $\Delta B$ = 0.045, are found through an exploration of parameter space about the best
fit values. Accounting for these $T$-dependences, the theoretical prediction of $I_{c0}^{\pm}(T) \propto
\alpha_{0}^{\pm}(T)M_{s}^{2}(T)(1 \pm B(T))/A(T)$ agrees with the measurement to within 10\% over the entire
range of $T$. All four devices that were extensively studied show an amplitude and $T$-dependence of
$\alpha_{0}$ very similar to that of Fig. 3(a); a gradual but significant increase with decreasing $T$ below 160
K, above which the devices are thermally unstable, followed by a stronger increase starting below 60 K - 40 K
where the best fit values of $\alpha_{0}$ also suggest differences between the two switching directions. For $T$
$<$ 60 K, the trends in the $T$-dependence of the spin torque parameters $A$ and $B$ vary from sample to sample,
but for $T$ $>$ 60 K both consistently show a very mild dependence on $T$ as illustrated in Fig. 3(b). For the
four samples studied in detail we found that at 40 K $A$ ranged from 0.5 to 0.68 and $B$ varied from 0.11 to
0.35. In general we also found that $A$ would decrease by 10 or 20\% in going from 40 K to 160 K while $B$ would
typically vary by 10\% or less.  These values of $A$ and $B$ and the variation with $T$ $>$ 60 K can be compared
with the results of a two-channel model~\cite{valet-fert-1993PRB} with which the measured GMR parameters, $R(T)$
and $\Delta R(T)$, can be used to predict the spin torque parameters~\cite{garcia-unpublished}. This model
predicts $A$ = 0.52, $B$ = 0.36 at 40 K, with $A$ decreasing to 0.47 at 160 K and $B$ remaining essentially
constant. This is in reasonable accord with the data, given the experimental uncertainties in nanomagnet size
and alignment.

\begin{figure}
\includegraphics[width=8.5cm]{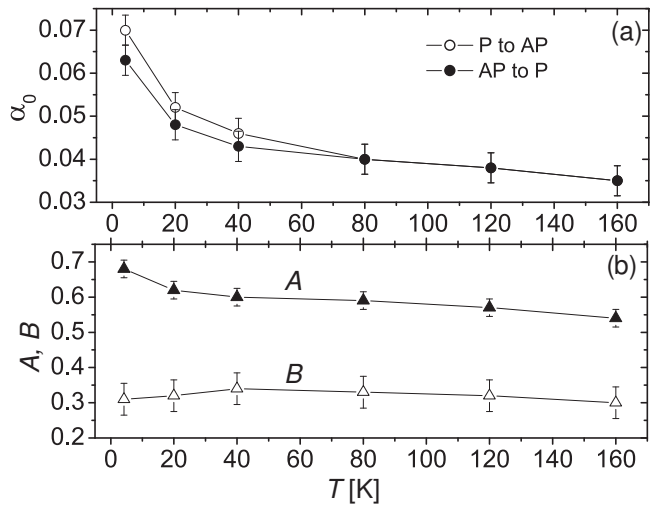}
\caption{Best fit parameters (a) damping $\alpha_{0}$ (for P to AP ($\circ$) and AP to P ($\bullet$) switching)
and (b) spin torque parameters $A$ ($\blacktriangle$) and $B$ ($\vartriangle$) as functions of $T$ from matching
simulated with measured values of 1/$t_{switch}$ versus $I$ for each $T$ for sample 2.}
\end{figure}

We attribute the significant $T$-dependence of $\alpha_{0}$ to the presence of a weak antiferromagnetic (AF)
layer on the sidewalls of the nanopillar.  Although no such AF layer was deliberately deposited, the exposure of
the nanopillars to air after ion mill definition undoubtedly oxidized the sidewalls, thus allowing for AF Py
oxide to form and weakly exchange bias the ferromagnetic layers. An example of direct evidence for this
adventitious exchange biasing is shown in Fig. 4, where switching fields $H_{S1}$ and $H_{S2}$, defined in the
inset, from 20 field scans at each $T$, are plotted from 4.2 K to 160 K for sample 3, an 80$\times$180 nm
ellipse, a previously unmeasured device cooled in $H_{\textrm{app}} = 0$. Note that $H_{S1}$ varies more rapidly
with $T$ than $H_{S2}$, which is indicative of an exchange bias that strengthens with decreasing $T$
(particularly rapidly below 40 K) and promotes AP alignment, i.e. a bias set by the dipole field from the fixed
layer.  Another key point illustrated in Fig. 4 is the large variation that develops in $H_{S1}$, and to a
lesser extent in $H_{S2}$, upon multiple minor loop scans after the device is cooled to low $T$. Initially, the
device switches repeatedly with nearly the same switching fields, but after six or seven magnetic reversals the
switching fields begin to fluctuate greatly from reversal to reversal, indicating stochastic variations in the
net strength of the oxide pinning field.  While the effects of the random pinning field are particularly
pronounced at 4.2 K they are observed up to 160 K, indicating that some degree of magnetic ordering within the
AF persists over this entire $T$ range and also that each reversal of the free layer nanomagnet has an
irreversible perturbing effect on the magnetic structure of the AF oxide. It is important to note that the slow
ramp rate current-driven switching events at low $T$ for these devices show good reproducibility, with little
variation from one sweep to another, as should be the case because ST switching currents are less sensitive to
field variations than are the switching fields.  The strength of this low $T$ AF exchange biasing varies from
device to device, with some samples showing no random variations in $H_{S,i}$. We do not believe that
fluctuations of this sort have affected any previously-published conclusions from our group. Nevertheless, the
fluctuations visible in some samples indicate clearly the presence of an AF layer that should influence the
properties of all nanopillar ST devices.

Exchange biasing in AF/Py films has been demonstrated to dissipate dynamic magnetic energy through a two-magnon
scattering process arising from local variations in the interfacial exchange
coupling~\cite{mcmichael-1998JAP,rezende-2001PRB,weber-2005APL}. Over the course of the pulsed $I$ measurements,
the free layer is switched hundreds of millions of times, which the $H_{S,i}$ data indicate should result in the
AF layer being on average magnetically ordered but with a finer, more randomized local magnetic structure that
leads to strong damping.  The rapid increase in damping, observed over the same low $T$ range where both the
unidirectional AF pinning field and the critical currents $I_{c0}^{\pm}(T)$ also increase rapidly, is attributed
to an increasing portion of the AF oxide layer becoming blocked, thereby simultaneously increasing the amount of
interfacial exchange coupling variation seen by the free layer nanomagnet as it moves in its dynamic switching
trajectory, consistent with the two-magnon model. The process of inducing randomization in the AF by the
nanomagnet reversal itself may also lead to enhanced damping. The unidirectional pinning field is present over
the entire $T$ range, with diminishing amplitude with increasing $T$. As the AF grains become unblocked, an
additional damping mechanism becomes possible if these grains can undergo reversal on the same time scale that
the free layer traces out its dynamic switching trajectory, which can result in domain drag or the ``slow
relaxer" dissipation process~\cite{mcmichael-2000JAP}. This effect could make a significant contribution to the
greater than intrinsic damping that persists to higher $T$.

\begin{figure}
\includegraphics[width=8.5cm]{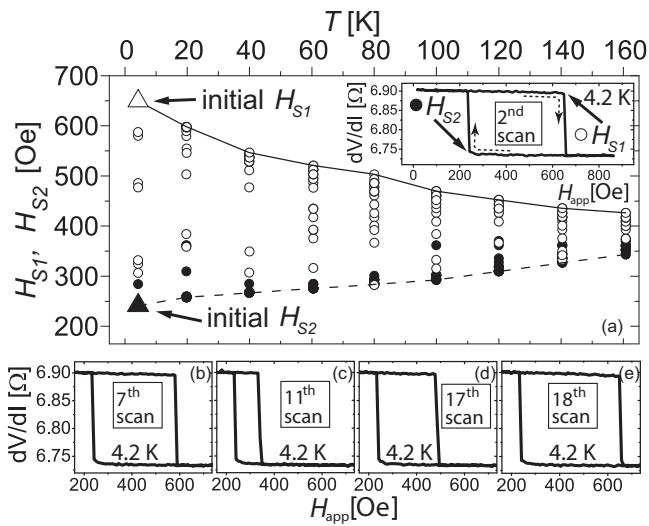}
\caption{(a) Switching fields $H_{S,i}$ of the free layer nanomagnet (defined in the inset) measured as a
function of $T$.  $H_{S1}$ ($\circ$) are AP to P and $H_{S2}$ ($\bullet$) are P to AP switching fields.
Progression of the random exchange field from the AF perimeter layer is observed in subsequent minor loop GMR
scans (b) through (e).  This previously unmeasured device (sample 3, an 80$\times$180 nm ellipse) was cooled to
4.2 K in $H_{\textrm{app}}$ = 0, whereupon a single, orientation-setting major loop scan, followed by 20 GMR
minor loop scans were taken. The sample was then sequentially warmed back to 160 K in 20 K steps, with 20 GMR
minor loops measured at each $T$. At 4.2 K, the free layer switched consistently at fields $H_{S1}$ $\approx$
650 Oe and $H_{S2}$ $\approx$ 240 Oe, shown as $\vartriangle$ and $\blacktriangle$ in (a), respectively, for the
first six GMR scans. Subsequent scans, however, showed more stochastic switching behavior that persisted for the
duration of the experiment. All 20 $H_{S,i}$ for each $T$ are shown, although some are indistinguishable due to
the size of the symbol.  The solid and dashed lines in (a) are guides to the eye for the maximal $H_{S1}$ and
$H_{S2}$ values, respectively, as functions of $T$.}
\end{figure}

Most ST device fabrication processes currently employed expose the sides of the free layer nanomagnet to some
level of an oxidizing ambient at some point, either during or after processing and to our knowledge there have
been no reports of actively protecting the sidewalls from oxidation.  We suggest that the native AF oxide layer
that forms can have substantial, previously under-appreciated consequences for the ST behavior, leading to a
substantially enhanced damping parameter which directly increases the critical currents for switching.  The
presence of this AF perimeter layer may also alter the boundary conditions that should be employed in
micromagnetic modeling of the free layer nanomagnet behavior and affect the dynamical modes of ST-driven
precession. We are currently investigating whether this AF perimeter layer can account, at least in part, for
the narrower than predicted ST-induced microwave oscillator linewidths that have been observed in similar
nanomagnets at low $T$~\cite{sankey-2005condmat}.

In summary, we have performed time-resolved measurements of the spin torque-driven switching of a Py nanomagnet
at $T$ = 4.2 K to 160 K.  LLG macrospin simulations are in close quantitative agreement with the ST switching
events, yielding values of the parameterized spin torque function $g(\theta) = A \sin\theta/(1 + B \cos\theta)$
and the damping parameter. We find $\alpha_{0}$ to be high, $>$ 0.03, and strongly $T$-dependent, which we
attribute to the AF pinning behavior of a thin Py oxide layer on the sidewall of the nanomagnet.   The values of
$A$ and $B$ are in fair numerical agreement with the spin torque calculated from the two-channel model using the
measured magnetoresistance values of the nanopillar spin-valve.  There is, however, considerable
device-to-device variation in the spin torque asymmetry parameter $B$, which we tentatively attribute to the
variable nature of the AF perimeter layer.  The presence of an AF oxide layer can have a major effect on the
nanomagnet dynamics. Controlling this layer will be important in optimizing spin torque-driven behavior.

This research was supported by ARO - DAAD19-01-1-0541, and by NSF through the NSEC support of the Cornell Center
for Nanoscale Systems.  Additional support was provided by NSF through use of the facilities of the Cornell
Nanoscale Facility - NNIN and the facilities of the Cornell MRSEC.

\bibliography{Switching_Speed_PRL}

\begin{thebibliography}{25}
\bibitem[{jit()}]{jitter}
\bibinfo{note}{Jitter noise in the triggering of the oscilloscope prevents the
  complete removal of distortions in the pulse shape coming from the pulser and
  amplifier, giving a slowly oscillating background and a shoulder in the
  current step. These introduce a maximum error of only 7\% in the measurements
  of $t_{switch}$ and are not significant. AP to P switching data (not shown)
  are very similar in both time and $I$ scales}.
\end{thebibliography}

\end{document}